\documentclass[prd,preprint,superscriptaddress,amsmath,amssymb,nofootinbib]{revtex4}
\usepackage{graphicx}
\usepackage{dcolumn}
\usepackage{bm}
\usepackage{amssymb}
\usepackage{amsmath}
\usepackage{epsfig}    
\usepackage{color}
\usepackage{slashed}
\usepackage{hhline}

\def\be{\begin{equation}}
\def\ee{\end{equation}}
\newcommand{\bea}{\begin{eqnarray}}
\newcommand{\eea}{\end{eqnarray}}
\newcommand{\nn}{\nonumber}

\begin{document}

\begin{flushright} {KIAS-P18117},  APCTP Pre2018 - 018 \end{flushright}

\title{Seesaw model with hidden $SU(2)_H \times U(1)_X$ gauge symmetry}

\author{Takaaki Nomura}
\email{nomura@kias.re.kr}
\affiliation{School of Physics, KIAS, Seoul 02455, Republic of Korea}

\author{Hiroshi Okada}
\email{hiroshi.okada@apctp.org}
\affiliation{Asia Pacific Center for Theoretical Physics (APCTP) - Headquarters San 31, Hyoja-dong,
Nam-gu, Pohang 790-784, Korea}
\affiliation{Department of Physics, Pohang University of Science and Technology, Pohang 37673, Republic of Korea}

\date{\today}

\begin{abstract}

We propose a seesaw model with a hidden gauge symmetry $SU(2)_H \times U(1)_X$ where two types of standard model singlet fermions in realizing a seesaw mechanism are organized into $SU(2)_H$ doublet. Then we formulate scalar and gauge sector, neutrino mass matrix and lepton flavor violations. In our gauge sector, $Z$-$Z'$ mixing appears after spontaneous symmetry breaking and we investigate constraint from $\rho$-parameter. In addition we discuss $Z'$ production at the large hadron collider via $Z$-$Z'$ mixing, where $Z'$ tends to dominantly decay into heavy neutrinos.

 \end{abstract}
\maketitle

\section{Introduction}

Generation of non-zero neutrino masses is one of the important issues which require an extension of the standard model (SM).
Moreover we expect smallness of the neutrino mass indicates a hint of structure of new physics beyond the SM.
Actually many mechanisms to generate neutrino masses are discussed such as canonical seesaw~\cite{Seesaw1, Seesaw2, Seesaw3, Seesaw4}, inverse seesaw~\cite{Mohapatra:1986bd, Wyler:1982dd}, linear seesaw mechanisms~\cite{Wyler:1982dd, Akhmedov:1995ip, Akhmedov:1995vm}, and so on. 
A linear seesaw mechanism (as well as inverse seesaw) is one of the interesting scenarios to realize tiny neutrino masses in which two types of SM singlet fermions are introduced; they are often denoted by $N_R^c$ and $S_L$.
In many cases, the introduction of these singlets are simply assumed in ad hoc way.
Even if we extend a gauge group such as left-right symmetry $SU(2)_L \times SU(2)_R \times U(1)_{B-L}$~\cite{Mohapatra:1986bd,Wyler:1982dd} only one type of singlet can be embedded in the right-handed lepton doublet.
Thus, a new hidden $SU(2)$ gauge symmetry is  one of the promising candidates to unify $N_R^c$ and $S_L$ in one doublet~\footnote{Another approach applying $SU(2)_L$ triplet fermion with hidden $U(1)$ symmetry can be referred to ref.~\cite{Nomura:2018ibs}.}.
 In order to have $SU(2)_H$ gauge anomaly cancellations for right-handed new fermions, even number of them is only allowed~\cite{Witten:1982fp}.
 This is also one of the unique natures of the $SU(2)_H$ gauge symmetry model, and we could obtain a specific feature such as prediction of one massless neutrino
  in the minimal scenario as we will discuss in the main text.

In this letter, we discuss a seesaw model with a hidden gauge symmetry of $SU(2)_H \times U(1)_X$ in which extra neutral fermions are introduced as $SU(2)_H$ doublet giving two types of SM singlet fermions after spontaneous symmetry breaking.
Introducing an $SU(2)_L \times SU(2)_H$ bi-doublet boson in our scalar sector, we can obtain Yukawa coupling among $SU(2)_L$ and $SU(2)_H$ lepton doublets which can realize the linear seesaw mechanism or Type-I seesaw like mechanism depending on parameter region.
Then we formulate neutrino mass matrix and lepton flavor violation (LFV) induced by the same Yukawa coupling generating the neutrino mass.
In addition,  we discuss  $Z$-$Z'$ mixing in  our gauge sector,
taking into account the constraint from $\rho$-parameter.
Finally we also consider collider physics in our model, focusing on $Z'$ production via the $Z$-$Z'$ mixing.
In our scenario, $Z'$ tends to dominantly decay into heavy neutrinos when it is kinematically allowed, and its branching ratio
shows clear difference from $Z'$ in other neutrino models with extra $U(1)$ such as $U(1)_{B-L}$ type as its $Z'$ should decay into SM fermions~\cite{Kang:2015uoc, Cox:2017eme, Accomando:2017qcs, Das:2017deo}.

This letter is organized as follows.
In Sec. II, we introduce our model, and formulate scalar sector, neutral gauge sector, neutrino mass matrix, and lepton flavor violations.
Then, we discuss collider phenomenologies focusing on $Z'$ boson which dominantly decays into heavy neutrinos.
 Finally we devote the summary of our results and the discussion in Sec.III.

\section{Model setup}
\begin{table}[t!]
\begin{tabular}{|c||c|c|c||c|c|c|c|c|}\hline\hline  
& ~$L_L^a$~& ~$e_R^a$~& ~$\Sigma_R^\alpha$~& ~$\Phi$~& ~$H_2$~& ~$H_1$~ & ~$\Delta$~ & ~$\varphi$~ \\\hline\hline 
$SU(2)_H$  & $\bm{1}$  &  $\bm{1}$ & $\bm{2}$ & $\bm{2}$ & $\bm{2}$ & $\bm{1}$ & $\bm{3}$ & $\bm{1}$   \\ \hline
$SU(2)_L$   & $\bm{2}$  & $\bm{1}$  & $\bm{1}$  & $\bm{2}$ & $\bm{1}$  & $\bm{2}$ & $\bm{1}$ & $\bm{1}$   \\\hline 
$U(1)_Y$    & $-\frac12$  & $-1$ & $0$  & $\frac12$  & $0$   &{$\frac12$} & $0$ & $0$ \\\hline
$U(1)_X$    & $0$  & $0$ & $0$ & $0$ & $x$ & $0$ & $0$ & $x$ \\ \hline   
\end{tabular}
\caption{Charge assignments of leptons and scalar fields including new field 
under $(SU(2)_H \times U(1)_X) \times (SU(2)_L \times U(1)_Y)$, where the upper index $a$ is the number of family that runs over 1-3
while $\alpha$ runs over $1$-$2n$  ($n$ is integer), and
all of them are singlet under $SU(3)_C$. }\label{tab:1}
\end{table}

In this section, we formulate our model in which we introduce hidden $SU(2)_H \times U(1)_X$ gauge symmetry.
In scalar sector, we introduce new scalar fields $H_2$,  $\Phi$, $\Delta$ and $\varphi$ which are doublet, doublet, real triplet and singlet under $SU(2)_H$ with $U(1)_X\times U(1)_Y$ charges $(x,0)$, $(0,1/2)$, $(0,0)$ and $(x,0)$, and only $\Phi$ is $SU(2)_L$ doublet while the others are singlet. 
Also SM-like Higgs doublet is denoted as $H_1$.
In our scenario, all these scalar fields develop vacuum expectation values (VEVs) inducing spontaneous symmetry breaking.
The scalar fields are written by their components as follows:
\begin{align}
& H_1 = \begin{pmatrix} h^+ \\ \frac{1}{\sqrt{2}} (v_1 + h_1^0 + i \eta_{h_1}) \end{pmatrix}, \quad H_2 = \begin{pmatrix} \varphi \\ \frac{1}{\sqrt{2}} (v_2 + h_2^0 + i \eta_{h_2}) \end{pmatrix}, \quad
\Delta = \frac{1}{\sqrt{2}} \begin{pmatrix} \delta_3 & \delta \\ \delta^* & - \delta_3 \end{pmatrix}, \nonumber \\
& \Phi = \begin{pmatrix} \phi_1^+ & \phi_2^+ \\ \frac{1}{\sqrt{2}} (\kappa_1 + \phi_1^0 + i \eta_{\phi_1}) & \frac{1}{\sqrt{2}} (\kappa_2 + \phi_2^0 + i \eta_{\phi_2}) \end{pmatrix}, \quad
 \varphi = \frac{1}{\sqrt{2}} (v_\varphi + \varphi_R + i \varphi_I),
\end{align}
where $v_{1,2}$ and $\kappa_{1,2}$ are VEVs for corresponding fields. 
The VEV of triplet is given by $\langle \delta_3 \rangle = v_\Delta/\sqrt{2}$ derived from scalar potential shown below.
In addition, $SU(2)_H$ doublet fermions $\Sigma_R^\alpha$ are introduced which is taken as right-handed and SM gauge singlet.
We write $\Sigma_R^\alpha$ with their components as
\begin{equation}
\Sigma_R^\alpha = \begin{pmatrix} N^\alpha_R \\ (S^{c }_L)^\alpha \end{pmatrix},
\end{equation}
where both component fields are electrically neutral, and $\alpha$ runs over $1$-$2n$ ($n$ is integer); we require even number of $\Sigma_R$ for guaranteeing the theory to be anomaly free~\cite{Witten:1982fp}.
In our discussion below, however, we fix $n$ to be 1 for simplicity: $\alpha = 1, 2$.

The mass term of $\Sigma_R$ and new Yukawa coupling are given by
\begin{align}
L =  \tilde M_{\alpha\beta} (\bar \Sigma_R)^\alpha (i \sigma_2) (\Sigma_R^c)^\beta - y_{\alpha\beta} \bar \Sigma_R^\alpha  \Delta (i \sigma_2) (\Sigma_R^c)^\beta  +  f_{\alpha\beta} \bar L^\alpha_L \tilde \Phi \Sigma_R^\beta  + h.c., \label{eq:Lfermion}
\end{align}
 where $\sigma_2$ is the second Pauli matrix and $\tilde \Phi \equiv (i \sigma^2) \Phi^* (i \sigma^2)$. 
 Note here that $\tilde M_{\alpha\beta}$ should be anti-symmetric matrix due to anti-symmetric contraction of $SU(2)_H$ indices in the term.
It suggests that $\tilde M$ reduces the matrix rank by one, and we cannot formulate the active neutrino mass matrix. Thus, we introduce $\Delta$ that leads to the second term as we will see later. The bi-doublet plays an role in inducing the Dirac mass that is also needed to construct the neutrino mass matrix.
 Moreover, $H_2$ and $\varphi$ play a role in breaking the gauge symmetry of $SU(2)_H\times U(1)_X$ spontaneously and avoiding massless Goldstone boson associated with breaking of global symmetry in the scalar potential.
Then scalar potential is written such as 
 \begin{align}
 \mathcal{V} = & - \tilde m_{H_1}^2 H_1^\dagger H_1 - \tilde m_{H_2}^2 H_2^\dagger H_2 - m_\varphi^2 \varphi^\dagger \varphi + \tilde m_\Delta^2 Tr[\Delta^\dagger \Delta] + \tilde m_\Phi^2 Tr[\Phi^\dagger \Phi] \nonumber \\
 & + \mu_\Delta (H_2^\dagger \Delta H_2 + h.c.) + \lambda   (\varphi^* H_1^\dagger \Phi H_2 +h.c.) + \lambda'  (\varphi H_1^\dagger \Phi \tilde H_2 +h.c.) + \lambda_\varphi (\varphi^*\varphi)^2  \nonumber \\
& + \lambda_{H_1} (H_1^\dagger H_1)^2 + \lambda_{H_2} (H_2^\dagger H_2)^2 
 + \lambda_\Phi Tr[\Phi^\dagger \Phi]^2 + \lambda_\Delta Tr[\Delta^\dagger \Delta]^2 + \lambda'_\Delta Tr[(\Delta^\dagger \Delta)^2]   
 \nonumber \\
& + \lambda_{H_1 H_2} (H_1^\dagger H_1)(H_2^\dagger H_2) + \lambda_{H_1 \Phi} (H_1^\dagger H_1)Tr[\Phi^\dagger \Phi] + \lambda_{H_2 \Phi} (H_2^\dagger H_2)Tr[\Phi^\dagger \Phi]  \nonumber \\
& + \lambda_{\Delta H_1} (H_1^\dagger H_1) Tr[\Delta^\dagger \Delta] + \lambda_{\Delta H_2} (H_2^\dagger H_2) Tr[\Delta^\dagger \Delta] +  \lambda_{\Delta \varphi} (\varphi^* \varphi) Tr[\Delta^\dagger \Delta] \nonumber \\
& + \lambda_{ H_1 \varphi} (H_1^\dagger H_1) (\varphi^* \varphi) + \lambda_{ H_2 \varphi} (H_2^\dagger H_2) (\varphi^* \varphi)  + \lambda_{ \Delta \Phi} Tr[\Delta^\dagger \Delta] Tr[\Phi^\dagger \Phi] \nonumber \\
& + \lambda_{  \Phi \varphi} (\varphi^* \varphi) Tr[\Phi^\dagger \Phi]  
+\lambda'_{\Delta H_2} \sum_{i=1}^3 (H^\dag_2\sigma_i H_2){\rm Tr}[\Delta^\dag\sigma^i\Delta]
+\lambda'_{\Delta \Phi} \sum_{i=1}^3{\rm Tr}[\Delta^\dag\sigma^i\Delta] (\Phi^\dag\sigma_i \Phi),
 \end{align}
 where $\tilde H_2 = i \sigma_2 H_2^*$ and we take all couplings as real parameters, and $\sigma_i$ (i=1,2,3) are Pauli matrices.
Note that the terms associated with operator $H_2^\dagger \Delta H_2$, $\varphi^* H_1^\dagger \Phi H_2$ and $\varphi H_1^\dagger \Phi \tilde H_2$ play a role to prevent massless Goldstone boson from appearing.
 Furthermore these terms can realize small VEVs of $\Phi$ and $\Delta$ which are preferred for neutrino mass generation.
 
 \subsection{Scalar sector}
 Firstly we assume $\varphi$ develops a VEV in higher scale compared to other VEV scale. 
 The VEV is derived by $\partial V/\partial v_\varphi =0$, providing $v_\varphi \simeq \sqrt{m_\varphi^2/\lambda_\varphi}$.
 Then the terms in mass parameter are modified as 
 \begin{align}
 \mathcal{V} \supset & - m_{H_1}^2 H_1^\dagger H_1 - m_{H_2}^2 H_2^\dagger H_2 + m_\Delta^2 Tr[\Delta^\dagger \Delta] + m_\Phi^2 Tr[\Phi^\dagger \Phi] \nonumber \\
 & + \mu (H_1^\dagger \Phi H_2 +h.c.) + \mu' ( H_1^\dagger \Phi \tilde H_2 +h.c.), \\
&  m_{X}^2 = \tilde m_{X}^2 - \lambda_{X\varphi} v_\varphi^2, \quad m_{Y}^2 = \tilde m_{Y}^2 + \lambda_{Y\varphi} v_\varphi^2, \quad \mu(\mu') = \lambda(\lambda') v_\varphi,
 \end{align} 
 where $X = \{H_1, H_2 \}$ and $Y = \{\Phi, \Delta\}$.
 The VEVs of the other scalar fields are obtained by solving the conditions
 \begin{equation}
\frac{\partial \mathcal{V}}{\partial v_1} = \frac{\partial \mathcal{V}}{\partial v_2} = \frac{\partial \mathcal{V}}{\partial \kappa_1} = \frac{\partial \mathcal{V}}{\partial \kappa_2} = \frac{\partial \mathcal{V}}{\partial v_\Delta} =  0.
\end{equation}
In our scenario, we require relations among VEVs such that $\kappa_{1,2} \ll v_{1,2}$ to realize a seesaw mechanism as discussed below.
Then VEVs are approximately given by
\begin{align}
& v_1 \simeq \sqrt{\frac{4 \lambda_{H_2} m_{H_1}^2 - 2 \lambda_{H_1 H_2} m_{H_2}^2}{4 \lambda_{H_1} \lambda_{H_2} - \lambda_{H_1 H_2}^2} }, \quad 
v_2 \simeq \sqrt{\frac{4 \lambda_{H_1} m_{H_2}^2 - 2 \lambda_{H_1 H_2} m_{H_1}^2}{4 \lambda_{H_1} \lambda_{H_2} - \lambda_{H_1 H_2}^2} }, \\
& \kappa_1 \simeq \frac{\sqrt{2} \mu' v_1 v_2}{2 m_{\Phi}^2 + \lambda_{H_1 \Phi}v_1^2 + \lambda_{H_2 \Phi} v_2^2 }, \quad 
\kappa_2 \simeq \frac{\sqrt{2} \mu v_1 v_2}{2 m_{\Phi}^2 + \lambda_{H_1 \Phi}v_1^2 + \lambda_{H_2 \Phi} v_2^2 }, \\
& v_\Delta \simeq \frac{\mu_\Delta}{2} \frac{v_2^2}{m_\Delta^2 },
\end{align}
where we chose $(\lambda_\Delta + \lambda'_{\Delta})v_\Delta^2 \ll m_\Delta^2$ and omit contribution from quartic terms assuming it is subdominant in deriving the triplet VEV; 
we also ignored $\lambda'_{\Delta H_2(\Phi)}$ coupling assuming it is sufficiently small for simplicity.
We thus see that $\kappa_{1,2}$ can be smaller than $v_{1,2}$ by choosing parameters $\mu$ and $\mu'$ to be small compared with other mass parameters.
In our case of $\kappa_{1,2} \ll v_{1,2}$ and assuming a mixing associated with $\{\varphi, \Delta\}$ is small, CP-even scalar bosons $h_1^0$ and $h_2^0$ from $H_1$ and $H_2$ can have sizable mixing.
Then squared mass matrix for $h_{1,2}^0$ is obtained as 
\begin{equation}
\mathcal{L} \supset \frac{1}{2} \begin{pmatrix}  h_1^0 \\ h_2^0 \end{pmatrix}^T \begin{pmatrix} 2\lambda_{H_1} v_1^2 & \lambda_{H_1 H_2} v_1 v_2 \\  \lambda_{H_1 H_2} v_1 v_2  & 2\lambda_{H_2} v_2^2 \end{pmatrix} \begin{pmatrix} h_1^0 \\  h_2^0 \end{pmatrix}.
\end{equation} 
The above squared mass matrix can be diagonalized, applying an orthogonal matrix that gives mass eigenvalues
\begin{equation}
m_{h,H}^2 = \lambda_{H_1} v_1^2 +\lambda_{H_2} v_2^2  \pm \sqrt{\left( \lambda_{H_1} v_1^2 -\lambda_{H_2} v_2^2 \right)^2 +  \lambda_{H_1 H_2}^2 v_1^2 v_2^2 }, 
\end{equation}
and the corresponding mass eigenstates $h$ and $H$ are obtained as   
\begin{equation}
\begin{pmatrix} h \\ H \end{pmatrix} = \begin{pmatrix} \cos \alpha & \sin \alpha \\ - \sin \alpha & \cos \alpha \end{pmatrix} \begin{pmatrix} h_1^0 \\ h_2^0 \end{pmatrix}, \quad
\tan 2 \alpha = \frac{ \lambda_{H_1 H_2} v_1 v_2}{\lambda_{H_1} v_1^2 - \lambda_{H_2} v_2^2},
\label{eq:scalar-mass-fields}
\end{equation}
where $\alpha$ is the mixing angle and $h$ is identified as the SM-like Higgs boson.

The mass eigenvalues for components of bi-doublet $\Phi$ are given by
\begin{align}
& m_{\phi_1^0}^2 = m_\Phi^2 + \frac{1}{2} \lambda_{H_1 \Phi} v_1^2 + \frac{1}{2} \lambda_{H_2 \Phi} v_2^2 + \lambda_\Phi (3 \kappa_1^2 + \kappa_2^2), \\
& m_{\phi_2^0}^2 = m_\Phi^2 + \frac{1}{2} \lambda_{H_1 \Phi} v_1^2 + \frac{1}{2} \lambda_{H_2 \Phi} v_2^2 + \lambda_\Phi ( \kappa_1^2 + 3\kappa_2^2), \\
& m_{\phi^\pm_{1,2}}^2 = m_{\eta_{\phi_{1,2}}}^2 = m_\Phi^2  + \frac{1}{2} \lambda_{H_1 \Phi} v_1^2 + \frac{1}{2} \lambda_{H_2 \Phi} v_2^2 + \lambda_\Phi (\kappa_1^2 + \kappa_2^2),
\end{align}
where corresponding components $\{ \phi_{1,2}^0, \phi_{1,2}^\pm, \eta_{\phi_{1,2} } \}$ can be approximately identified with mass eigenstates for small $\kappa_{1,2}$.
In addition, the mass eigenvalues are almost degenerated in our case.
 Scalar bosons from $\Delta$ are neutral scalar bosons and it would interact with SM particle via neutral fermion mixing and Higgs mixing. 
In this paper we just assume $\Delta$ is heavy whose mass is dominantly given by $m_\Delta$. 
 
 \subsection{Gauge sector}
 
\begin{figure}[t]
\begin{center}
\includegraphics[width=70mm]{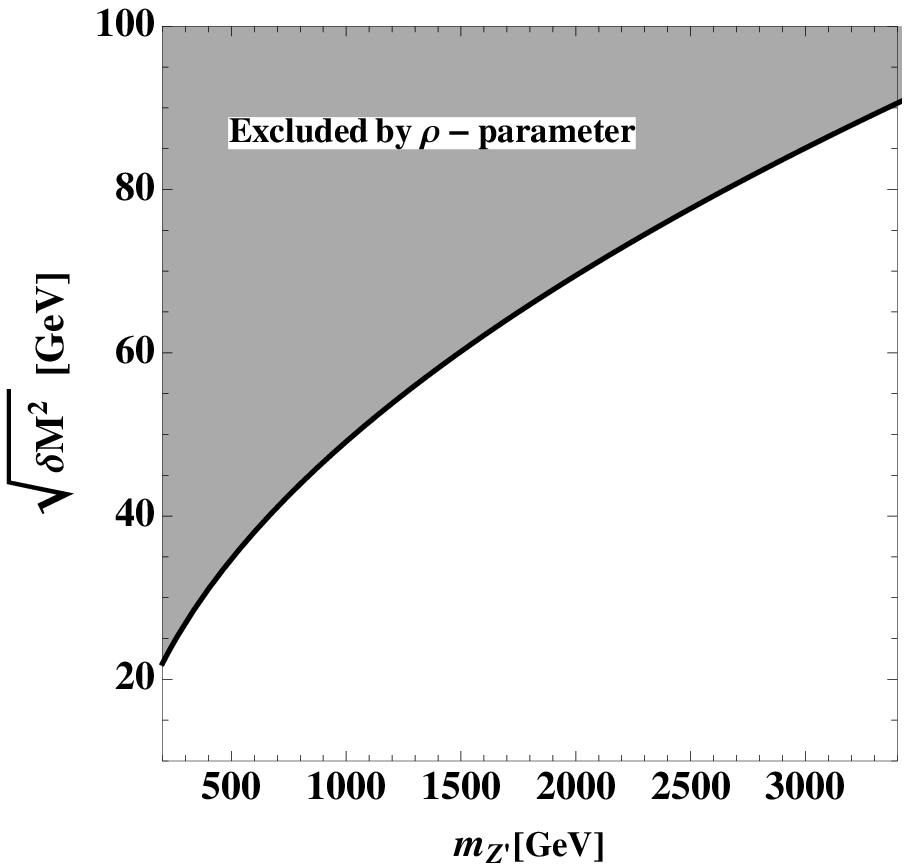}
\includegraphics[width=70mm]{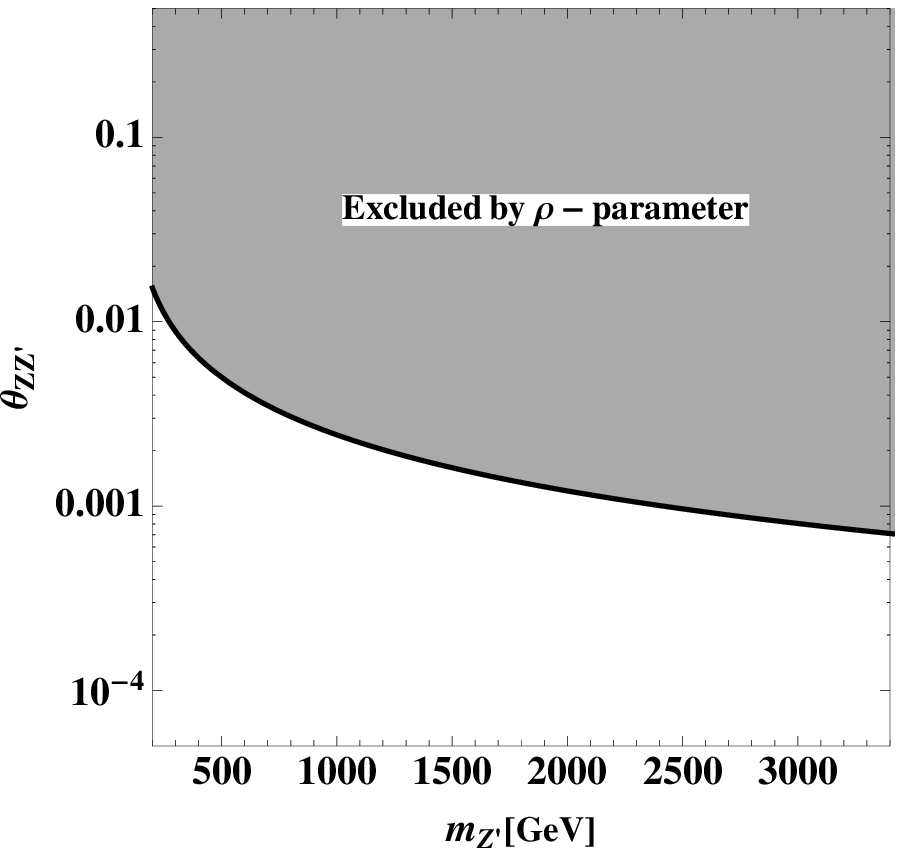} 
\caption{Left: Solid line indicate the upper limit of $\sqrt{\delta M^2}$ allowed by $\rho$-parameter constraint as a function of $m_{Z'}$. Right: Solid line indicate the corresponding upper limit of $\theta_{ZZ'}$ as 
a function of $m_{Z'}$.} 
  \label{fig:constraints}
\end{center}\end{figure}

 Here we analyze mass terms for gauge fields. 
 The mass terms are obtained after spontaneous breaking of $SU(2)_H \times SU(2)_L \times U(1)_Y \times U(1)_X$ gauge symmetry via kinetic terms of scalar fields:
 \begin{align}
 L_{K} = & (D_\mu H_1)^\dagger (D^\mu H_1) + (D_\mu H_2)^\dagger (D^\mu H_2) + Tr[(D_\mu \Phi)^\dagger (D^\mu \Phi)] +  (D_\mu \varphi)^\dagger (D^\mu \varphi),  \\
 D_\mu \Phi = & \partial_\mu \Phi - i g_2 W_\mu^i \frac{\sigma^i}{2} \Phi + i g_H \frac{\sigma^i}{2} \Phi W_{H_\mu}^i - i g_1 \frac{1}{2} B_\mu \Phi, \\
 D_\mu H_1 = & \partial_\mu H_1 - i g_2 W_\mu^i \frac{\sigma^i}{2} H_2 - i g_1 \frac{1}{2} B_\mu H_1, \\
 D_\mu H_2 = & \partial_\mu H_2 - i g_H \frac{\sigma^i}{2} W_{H_\mu}^i H_2 - i x g_X X_\mu H_2, \\
 D_\mu \Delta = & \partial_\mu \Delta -i g_H \left[ \frac{\sigma^i}{2} W_{H_\mu}^i, \Delta \right], \\
  D_\mu \varphi = &  \partial_\mu \varphi - i x g_X X_\mu \varphi,
 \end{align}
 where $\sigma^i$ denotes the Pauli matrix, $W_{H_\mu}^i$ and $X_\mu$ are $SU(2)_H$ and $U(1)_X$ gauge fields, 
 and $g_{1}$, $g_2$, $g_H$ and $g_X$ are respectively gauge couplings for $U(1)_Y$, $SU(2)_L$, $SU(2)_H$ and $U(1)_X$.
 Then the mass terms for gauge fields are given by
 \begin{align}
 L_M = & \frac{1}{8} \bigl[ (g_1^2 + g_2^2) (v_1^2 + \kappa^2) \tilde Z_\mu \tilde Z^\mu + 2 g_2^2 (v_1^2 + \kappa^2) W^+_\mu W^{- \mu}  \nonumber \\
& \quad + g_H^2 (v_2^2 + \kappa^2 + v_\Delta^2 )(W_{H}^{1\mu} W_{H \mu}^{1} + W_{H}^{2\mu} W_{H \mu}^{2} ) 
+  v_2^2 (g_H W_{H \mu}^3 - 4 x g_X X_\mu)^2  \nonumber \\
& \quad + 2 g_H \sqrt{g_1^2 + g_2^2} \tilde Z_\mu (\Delta \kappa^2 W_H^{3 \mu}  + 2 \kappa_1 \kappa_2 W_H^{1\mu}) \bigr],
 \end{align}
 where  we define
 \begin{equation}
 \kappa^2 = \kappa_1^2 + \kappa_2^2, \quad \Delta \kappa^2 = \kappa_1^2 - \kappa_2^2, \quad \tilde Z_\mu = \frac{1}{\sqrt{g_1^2 + g_2^2}} (g_1 B_\mu - g_2 W_\mu^3).
 \end{equation}
 For $W^\pm$ boson, the mass is given by 
 \begin{equation}
 m_W = \frac12 g_2 \sqrt{v_1^2 + \kappa^2},
 \label{eq:Gmass1}
 \end{equation}
 where $\sqrt{v_1^2 + \kappa^2} = v \simeq 246$ GeV is required. 
 In our following, analysis we take $\kappa_1 \sim \kappa_2$ so that the mass term associated with $\Delta \kappa^2$ is negligibly small compared to other mass terms.
 We also do not discuss $W^3_\mu$-$X_\mu$ mixing, since it does not couple with SM sector directly and focus on $W^1_\mu$-$\tilde Z_\mu$ sector.
 Then $W_{H \mu}^{1}$ mainly mixes with $\tilde Z_\mu$ and corresponding mass matrix is given by
 \begin{align}
&  L_M \supset \frac{1}{2} \begin{pmatrix} \tilde Z_\mu \\ W_{H \mu}^{1} \end{pmatrix}^T 
 \begin{pmatrix} M_{\tilde Z}^2 & \delta M^2 \\ \delta M^2 & M_{X}^2 \end{pmatrix}  
 \begin{pmatrix} \tilde Z_\mu \\ W_{H \mu}^{1} \end{pmatrix}, \\
& M_{\tilde Z}^2 = \frac{1}{4} (g_1^2 + g_2^2) (v_1^2 + \kappa^2), \quad M_{X}^2 = \frac{1}{4} g_H^2 (v_2^2 + \kappa^2), \quad \delta M^2 = \frac{1}{2} g_H \sqrt{g_1^2+g_2^2} \kappa_1 \kappa_2.
\label{eq:Gmass2}
 \end{align}
 Diagonalizing the mass matrix, we obtain mass eigenvalues 
 \begin{equation}
 m_{Z,Z'}^2 = \frac{M_{\tilde Z}^2 + M_X^2}{2} \pm \frac{\sqrt{(M_{\tilde Z}^2 - M_X^2)^2 + 4 \delta M^4}}{2}, 
 \end{equation}
 and mass eigenstates are given by
 \begin{align}
& \begin{pmatrix} Z_\mu \\ Z'_\mu \end{pmatrix} = \begin{pmatrix} \cos \theta_{ZZ'} & \sin \theta_{ZZ'} \\ - \sin \theta_{ZZ'} & \cos \theta_{ZZ'} \end{pmatrix} \begin{pmatrix} \tilde Z_\mu \\ W^1_{H \mu} \end{pmatrix}, 
\label{eq:mixing} \\
& \sin 2 \theta_{ZZ'} = \frac{2 \delta M^2}{m_{Z}^2 - m_{Z'}^2}. \label{eq:angle}
 \end{align}
Here we consider the limit of $M_{\tilde Z}^2, \delta M^2 \ll M_X^2$ and mass eigenvalues are approximately 
\begin{equation}
m_{Z}^2 \simeq M_{\tilde Z}^2 - \frac{\delta M^4}{M_X^2}, \quad m_{Z'}^2 \simeq M_X^2 + \frac{\delta M^4}{M_X^2},
\end{equation}
where $m_Z$ is identified as the SM Z boson mass.
Thus $\rho$-parameter in the model is shifted from 1 and given as 
\begin{equation}
\rho \equiv \frac{m_W }{m_Z \cos \theta_W} = \frac{M_{\tilde Z}^2}{m_Z^2} \simeq 1 + \frac{\delta M^4}{m_Z^2 m_{Z'}^2},
\end{equation}
where we used Eqs.~(\ref{eq:Gmass1}) and (\ref{eq:Gmass2}) to obtain relation between $m_W$ and $M_{\tilde Z}$.
Then we obtain allowed parameter region on $\{\delta M^2, m_{Z'}\}$ space from observed $\rho$-parameter $\rho = 1.0004^{+0.0003}_{-0.0004}$~\cite{PDG} with $2 \sigma$ error.
In the left plot of Fig.~\ref{fig:constraints}, we indicate the upper limit of $\sqrt{\delta M^2}$ as a function of $m_{Z'}$, while corresponding upper limit of $\theta_{ZZ'}$ is given in the right plot.
We thus find that the VEVs in bi-doublet scalar are required not to be large, assuming gauge coupling $g_H$ is $\mathcal{O}(0.1)$ to $\mathcal{O}(1)$.

 \subsection{Neutral fermion mass}
Here we consider neutral fermion masses including active neutrino masses.
Firstly mass term for $\Sigma_R^\alpha$ can be written in component form:
\begin{align}
& \tilde M_{\alpha\beta} (\bar \Sigma_R)^\alpha (i \sigma_2) (\Sigma^c)^\beta_R + y_{\alpha\beta} \bar \Sigma_R^\alpha \langle \Delta \rangle (i \sigma_2) (\Sigma_R^c)^\beta  \nonumber \\
& =   \tilde M_{\alpha\beta} \left[ (\bar N_R)^\alpha (S_L)^\beta - (\bar S_L^c)^\alpha (N_R^c)^\beta \right] + \frac{y_{\alpha \beta} v_\Delta}{2} \left[ (\bar N_R)^\alpha (S_L)^\beta + (\bar S_L^c)^\alpha (N_R^c)^\beta \right] \nonumber \\
& \equiv M_{\alpha\beta} (\bar N_R)^\alpha (S_L)^\beta,
\label{eq:massM}
\end{align} 
where $M_{\alpha\beta}$ is general $2 \times 2$ mass matrix. Note that we don't have diagonal term of $M_{\alpha \beta}$ without VEV of $\Delta$.
 After $\Phi$ developing VEV, we obtain mass terms from the Lagrangian in Eq.~(\ref{eq:Lfermion}) such that 
\begin{equation}
L \supset  \frac{ f_{a\beta} \kappa_1}{\sqrt{2}} \bar \nu_L^a (S_L^c)^\beta - \frac{f_{a\beta} \kappa_2}{\sqrt{2}} \bar \nu_L^a N_R^\beta + h.c. \, ,
\end{equation}
where $a = 1$-$3$ and $\beta = 1,2$.
The mass matrix for neutral fermion is then obtained as\footnote{$U(1)_X$ forbids $H^T_2(i\sigma_2)\Delta H_2$ that leads to the non-vanishing components of (22) and (33) in the neutral fermion mass matrix, since it develops nonzero VEV of $\Delta$. Therefore, our model would spoil without $U(1)_X$.} 
\begin{equation}
L_{mass} = \begin{pmatrix} \bar \nu^c_L \\ \bar N_R \\ \bar S^c_L \end{pmatrix}^T
\begin{pmatrix} 0 & M_{\kappa_2}^* & M_{\kappa_1}^* \\ M_{\kappa_2}^\dagger & 0 & M \\ M_{\kappa_1}^\dagger & M^T & 0 \end{pmatrix}
\begin{pmatrix}  \nu_L \\  N_R^c \\  S_L \end{pmatrix},
\end{equation}
where $(M_{\kappa_{1,2}})_{a \beta} = f_{a \beta} \kappa_{1,2}/\sqrt{2}$ and $M$ is given by Eq.~(\ref{eq:massM}).
Here we assume $M_{\kappa_{1,2}} \ll M$ in our scenario and following situations can be considered depending on relative size of $\kappa_1$ and $\kappa_2$:
\begin{itemize}
\item $\kappa_1 \simeq \kappa_2$ and we obtain mass matrix similar to type-I seesaw mechanism. 
\item $\kappa_2 \ll \kappa_1$ and we obtain linear seesaw like hierarchy for the components in the mass matrix.
\end{itemize} 
In our analysis, we take $\kappa_1 \simeq \kappa_2 \simeq \kappa/\sqrt{2}$ for simplicity and define $(M_{NS})_{a\beta} \equiv f_{a\beta} \kappa/2$.
In fact it is more natural case since there is no reason to have $\lambda \ll \lambda'$ for generation hierarchy between $\kappa_1$ and $\kappa_2$.
As a result, mass matrix for neutral fermions can be obtained as 
\begin{equation}
L_{mass} = \begin{pmatrix} \bar \nu^c_L \\ \bar N_R \\ \bar S^c_L \end{pmatrix}^T
\begin{pmatrix} 0 & M_{NS}^* & M_{NS}^* \\ M_{NS}^\dagger & 0 & M \\ M_{NS}^\dagger & M^T & 0 \end{pmatrix}
\begin{pmatrix}  \nu_L \\  N_R^c \\  S_L \end{pmatrix}.
\end{equation}
Applying seesaw approximation with $M_{NS} \ll M$, we then obtain active neutrino mass such that
\begin{equation}
- m_\nu \approx   M_{NS}^*[M^{-1} +(M^T)^{-1} ] M_{NS}^\dagger   \equiv M_{NS}^*M_S^{-1} M_{NS}^\dagger=
 M_{NS}^* R^{-1} (R^T)^{-1} M_{NS}^\dagger.
\end{equation}
Note here that $M_S$ is uniquely decomposed by  a lower unit triangular matrix $R^{}$, since $M_S$ is the symmetric matrix~\cite{Nomura:2016run}.
Then $M_{NS}$ is rewritten in terms of experimental values as
\begin{equation}
M_{NS}^* =  i U^T \sqrt{D_\nu} O_{} R,
\end{equation}
where $O$ is an arbitrary three by two matrix with $O^TO=1_{2\times2}$ and $OO^T={\rm diag}(0,1,1)$, $m_\nu\equiv U^T D_\nu U$, $D_\nu$ is mass eigenvalues of neutrinos, and $U$ is the unitary matrix to diagonalize the neutrino mass matrix.
Note here that in our scenario, we predict one massless neutrino in which we assume minimal number of $SU(2)_H$ doublet chiral fermion for anomaly cancellation.
Next, we have to consider the constraint from non-unitarity, and this can be evaluated by $|\epsilon|\equiv \delta_N\delta_N^\dag$~\cite{Das:2012ze, Das:2017nvm, Das:2017ski};
\begin{align}
|\epsilon|\approx
\begin{pmatrix} 0.006\pm0.0063 &<1.29\times10^{-5} & <8.76\times10^{-3} \\ <1.29\times10^{-5} & 0.005\pm0.0063 & <1.05\times10^{-2} \\ 
<8.76\times10^{-3} & <1.05\times10^{-2} & 0.005\pm0.0063 \end{pmatrix},
\end{align}
where $\delta_N \equiv M_{NS}^* M^{-1}$ and $\delta_N <<1$ is expected. 
Note that condition $M_{NS} \ll M$ can be easily achieved by taking VEV of bi-doublet to be small which is also motivated by $\rho$-parameter constraint discussed above.
Rough estimation leads to $|\epsilon|\approx |D_\nu/M_{NS}|^2$, and this should conservatively satisfy $|\epsilon|\lesssim 10^{-5}$.
Therefore, we find
\begin{align}
 32\ {\rm eV}\lesssim M_{NS}, 
 \end{align}
 where we fix to be $D_\nu\sim$0.1 eV. 
 In fact, required order of $M_{NS}$ is roughly $M_{NS} \sim \sqrt{D_\nu M} \sim 10^{-4}$ GeV for $M \sim 100$ GeV which satisfies the condition above.
Heavier fermions are also diagonalized by the unitary matrix and their mass eigenvalues are degenerately given by $M_{N_{1,2}}\approx\frac{M+M^T}{2}$ and their eigenstates are found to be   
\begin{align}
\begin{pmatrix}  N_R^c \\ \bar S_L \end{pmatrix}\approx
\begin{pmatrix}   \frac{1}{\sqrt2} & - \frac{i}{\sqrt2} \\  \frac{1}{\sqrt2}  &  \frac{i}{\sqrt2}  \end{pmatrix}
\begin{pmatrix}    N_1 \\  N_2 \end{pmatrix}_L,
\end{align}
where index for generation is omitted here.

 \subsection{Yukawa interactions and lepton flavor violation}
 The Yukawa interactions including SM charged leptons are obtained from third term of Eq.~(\ref{eq:Lfermion}) such that
 \begin{align}
 f_{a \beta} \bar L^a_L \tilde \Phi \Sigma_R^\beta + h.c. &\supset f_{a\beta} \left[ \bar \ell^a_L N_R^\beta \phi_2^- - \bar \ell_L^a (S_L^c)^\beta \phi_1^- \right] + h.c.
\nn\\
&\approx
\frac{ f_{a \beta}}{\sqrt2} \left[ \bar \ell^a P_R (N_1^\beta-iN_2^\beta) \phi_2^- - \bar \ell^a P_R(N_1^\beta+iN_2^\beta) \phi_1^- \right] + h.c.
 \end{align}
 where $\{\ell^1, \ell^2, \ell^3\} =\{e, \mu, \tau\}$ and we omit interactions containing only neutral fermions.
 Then the formula of lepton flavor violations (LFVs), $\ell_a \to \ell_b \gamma$,  is given by~\cite{Lindner:2016bgg, Baek:2016kud}
\begin{align}
 {\rm BR}(\ell_a\to\ell_b\gamma)\approx \frac{48\pi^3\alpha_{\rm em} C_{ab} }{{\rm G_F^2} } 
\left|a_{R_{ab}}(N_1^k,\phi_1^-)+a_{R_{ab}}(N_2^k,\phi_1^-)+a_{R_{ab}}(N_1^k,\phi_2^-)+a_{R_{ab}}(N_2^k,\phi_2^-)\right|^2,
 \end{align}
where $C_{21}\approx$1, $C_{31}\approx$0.1784, $C_{32}\approx$0.1736, $G_F\approx1.17\times10^{-5}$ GeV$^{-2}$, and
 \begin{align}
 a_{R_{ab}}(\rho,\sigma) &\approx \frac{1}{ {2(4\pi)^2}}
 \sum_{k=1}^2 {f_{bk} f^\dag_{ka}}
 \int_0^1dx\int_0^{1-x}dy\frac{xy}{(x^2-x)m^2_{\ell_a} +x m_\rho^2+(1-x) m^2_\sigma}.
\end{align} 
Experimental upper bounds for these LFV processes are respectively given by ${\rm BR}(\mu\to e\gamma)\lesssim 4.2\times10^{-13}$, ${\rm BR}(\tau\to e\gamma)\lesssim 3.3\times10^{-8}$, and ${\rm BR}(\tau\to \mu\gamma)\lesssim 4.4\times10^{-8}$~\cite{TheMEG:2016wtm, Adam:2013mnn}.
 We find that the LFV constraints can be easily avoided. For example, taking $m_{\phi_{1,2}^\pm} = 1000$ GeV and $m_{N_{1,2}^k} = 400$ GeV, 
current $\mu \to e \gamma$ constraint of $BR(\mu \to e \gamma) < 4.2 \times 10^{-3}$ require Yukawa couplings to satisfy $\sum_{k=1}^2 {f_{bk} f^\dag_{ka}} \lesssim 0.1$.

 \subsection{Collider physics}

\begin{figure}[t]
\begin{center}
\includegraphics[width=70mm]{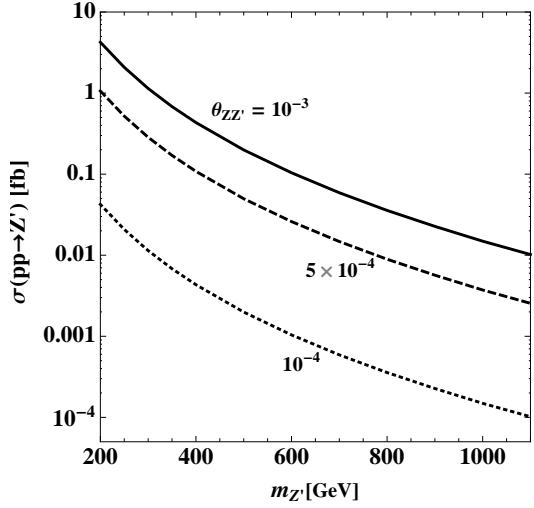}
\caption{$Z'$ production cross section at the LHC 14 TeV as a function of $m_{Z'}$ for several values of $\theta_{ZZ'}$.} 
  \label{fig:CXZp}
\end{center}\end{figure}

Here we discuss $Z'$ production at the LHC.
In our model, $Z'$ can be produced via $Z-Z'$ mixing where interaction among $Z'$ and the SM fermions is obtained as:
\begin{equation}
\mathcal{L} \supset  g_2 \sin \theta_{Z Z'} Z'_\mu J_Z^\mu, 
\end{equation}
where $J_Z^\mu$ is the neutral current in the SM. 
Then the $Z'$ production cross section via Drell-Yang process is proportional to suppression factor of $\sin^2 \theta_{ZZ'}$.   
Here we estimate $Z'$ production cross section using {\it CalcHEP}~\cite{Belyaev:2012qa} by use of the CTEQ6 parton distribution functions (PDFs)~\cite{Nadolsky:2008zw}, implementing relevant interactions.
In  Fig.~\ref{fig:CXZp}, we show $Z'$ production cross section at the LHC 14 TeV as a function of $m_{Z'}$ for several values of $\theta_{ZZ'}$.
We thus find that $\theta_{ZZ'} \gtrsim 10^{-4}$ is preferred to obtain the cross section which could be tested at the LHC.
Note also that $\kappa$ cannot be too small to obtain sizable $\theta_{ZZ'}$ value. From Eq.~(\ref{eq:angle}), we estimate 
\begin{equation}
\theta_{ZZ'} \sim 0.18 g_H \frac{\kappa^2}{m_{Z'}^2} \sim 3 \times 10^{-4} g_H \left( \frac{\kappa}{20 \ {\rm GeV}} \right)^2 \left( \frac{500 \ {\rm GeV}}{m_{Z'}} \right)^2,
\end{equation}
where we used $m_{Z}^2 \ll m_{Z'}^2$.
Thus we should require $\kappa \gtrsim 10$ GeV to obtain $\theta_{ZZ'} \gtrsim 10^{-4}$ assuming $g_H$ is $\mathcal{O}(1)$ value. 
In that case Yukawa coupling $f_{a \beta}$ is $ \sim \mathcal{O}(10^{-5})$ to realize $M_{NS} \sim 10^{-4}$ GeV for neutrino mass generation.

In our model, $Z'$ dominantly decays into extra neutral fermions $N_{1,2}$, if the decay process is kinematically allowed 
where $Z' \bar N_i N_i$ terms are obtained as: 
\begin{align}
\mathcal{L} & \supset  \bar \Sigma_R^\alpha D_\mu \gamma^\mu \Sigma_R^\alpha \nonumber \\
& \supset g_H \cos \theta_{ZZ'} Z'^\mu (\bar N_2^\alpha \gamma_\mu \gamma_5 N_2^\alpha - \bar N_1^\alpha \gamma_\mu \gamma_5 N_1^\alpha), 
\end{align}
where we have applied Eq.~(\ref{eq:mixing}).
Note that $Z'$ can also decay into scalar bosons from bi-doublet but we assume these scalar bosons are heavy and the decay modes are kinematically forbidden.
Branching ratios (BRs) of $Z' \to \bar f_{SM} f_{SM}$ ($f_{SM}$ denotes a SM fermion) are suppressed by small $\sin \theta_{ZZ'}$, and we have
$BR(Z' \to \bar f_{SM} f_{SM})/BR(Z' \to \bar N_i N_i) \propto \sin^2 \theta_{ZZ'} g_2^2/g_H^2$.
Thus one finds $BR(Z' \to \bar f_{SM} f_{SM}) \ll BR(Z' \to \bar N_i N_i)$, if gauge coupling $g_{H}$ is not too small.
Then collider constraints from $pp \to Z' \to \bar f_{SM} f_{SM}$ are not significant in our model, requiring $g_H \gg \sin \theta_{ZZ'}/g_2 $. 
Here we assume that the mass of $Z'$ satisfies $2 m_{N_{1,2}^1} < m_{Z'} < 2 m_{N_{1,2}^{2}}$ so that $Z'$ decays into $\bar N^1_{1(2)} N^1_{1(2)}$ pair; $N^1_i$ denote the lighter mass eigenstate and we omit the upper index in the following. 
Then $N_{1,2}$ decays as $N_{1,2} \to \ell^\pm W^\mp, Z \nu, h \nu$ via light-heavy neutrino mixing.
In Table.~\ref{tab:CX}, we show $\sigma(pp \to Z' \to \bar N_{1,2} N_{1,2} \to W^\pm W^\pm \ell^\mp \ell^\mp)$ at the LHC 14 TeV for some benchmark values of $(m_{Z'}, \sin \theta_{ZZ'})$, adopting $BR(N \to W^\pm \ell^\pm) \simeq 0.5$, where we assume mass of $N_{1,2}$ is sufficiently lighter than $m_{Z'}/2$.
We find that cross section of $\sim 0.22$ fb can be obtained if $m_{Z'}$ is 400 GeV and $\sin \theta_{ZZ'} = 10^{-3}$, which could be tested at the future LHC experiments with integrated luminosity of 300 fb$^{-1}$.
The signals of the process are multi-lepton final state and the same sign charged lepton plus jets depending on decay mode of $W$ boson. 
The cross section of relevant SM background processes are estimated as 
\begin{equation}
\sigma(pp \to W^+ W^- \ell^+ \ell^-) \simeq 5.1 \ {\rm fb}, \ \sigma(pp \to W^+ Z Z ) \sim 20 \ {\rm fb}, \ \sigma(pp \to W^- Z Z ) \sim 10 \ {\rm fb}, 
\end{equation}
where these processes provide multi-lepton and jets via decay of W and Z bosons.
In fact, cross section of these SM processes are larger than our signal cross section and we need relevant selection and kinematical cuts to suppress number of background events.
More detailed analysis including detector simulation and cut analysis is beyond the scope of this paper. 
Also if $Z'$ is heavier and/or $\sin \theta_{ZZ'}$ is smaller, the cross section becomes much smaller but it could be accessible at the high-luminosity (HL) LHC with integrated luminosity of 3000 fb$^{-1}$.
Note that we can distinguish our $Z'$ from other $Z'$ such as that from $U(1)_{B-L}$, since $Z' \to \bar f_{SM} f_{SM}$ mode is expected to be absent in our case.

\begin{table}[t]
\begin{tabular}{|c||c|c|c|}\hline\hline  
$(m_{Z'}, \sin \theta_{ZZ'})$ & $(400 \ {\rm GeV}, 10^{-3})$ & $(800 \ {\rm GeV}, 10^{-3})$ & $(400 \ {\rm GeV}, 10^{-4})$ \\ \hline
$\sigma BR$ & 0.22 fb & 0.018 fb & 0.0022 fb \\ \hline
\end{tabular}
\caption{$\sigma(pp \to Z' \to \bar N_{1,2} N_{1,2})BR(N_{1,2} \to W^\pm W^\pm \ell^\mp \ell^\mp)$ at the LHC 14 TeV for some benchmark values of $(m_{Z'}, \sin \theta_{ZZ'})$ }\label{tab:CX}
\end{table}

Before closing this section, we discuss the width of heavy neutrino $N_{1,2}$ in our scenario.
Parametrizing mixing between active neutrino and heavy neutrino by $\theta_{\nu N}$, we estimate the decay width for $N_{1,2} \to W^\pm \ell^\mp$ by
\begin{equation}
\Gamma_{N \to W^\pm \ell^\mp} \simeq \frac{g_2^2 \theta^2_{\nu N}}{64 \pi} \frac{(m_N^2 - m_W^2)^2}{m_N^3} \left( 2 + \frac{m_N^2}{m^2_W} \right),
\end{equation}
where $m_N$ indicate heavy neutrino mass.
Taking $m_N = 150$ GeV we obtain $(\Gamma_{N \to W^\pm \ell^\mp})^{-1} \sim 1.8 \times 10^{-16} \theta_{\nu N}^{-2} [{\rm m}]$ for decay length. 
Thus heavy neutrino decays before reaching detector for $\theta_{\nu N} \sim M_{NS}/m_N \sim 10^{-6}$.

 \section{Summary and discussions}
 
 In this paper we have proposed a seesaw model based on a hidden gauge symmetry $SU(2)_H \times U(1)_H$ in which two types of singlet fermions to realize a seesaw mechanism 
 are unified into a doublet of hidden $SU(2)_H$. Then a Yukawa interaction among the $SU(2)_H$ doublet fermion and the SM lepton doublet  is realized by introducing bi-doublet scalar filed under $SU(2)_L \times SU(2)_H$. 
 
Then we have  formulated scalar sector and gauge sector of our model taking into account $\rho$-parameter constraint from $Z$-$Z'$ mixing.
The neutral fermion mass matrix has been analyzed in which active neutrino mass is derived via Type-I like seesaw mechanism.
In our scenario, we predict one massless neutrino in which we assume minimal number of $SU(2)_H$ doublet chiral fermion for anomaly cancellation.
We have also taken into account constraints from non-unitarity and LFV, and found the constraints can be avoided easily.

Finally we have discussed collider physics, focusing on $Z'$ production via $Z$-$Z'$ mixing. 
Our $Z'$ can dominantly decay into heavy neutrinos $N_{1,2}$ and a SM fermion pair decay mode tends to be absent due to suppression by small $Z$-$Z'$ mixing effect.
Then cross section of $\sim 0.22$ fb can be obtained for $pp \to Z' \to \bar N_{1,2} N_{1,2} \to W^\pm W^\pm \ell^\mp \ell^\mp$ with $(m_{Z}, \sin \theta_{ZZ'}) = (400 \ {\rm GeV}, 10^{-3})$ 
which would be tested by future LHC experiments. 
More parameter region can be tested at the HL-LHC.

\section*{Acknowledgments}
This research was supported by an appointment to the JRG Program at the APCTP through the Science and Technology Promotion Fund and Lottery Fund of the Korean Government. This was also supported by the Korean Local Governments - Gyeongsangbuk-do Province and Pohang City (H.O.). H. O. is sincerely grateful for the KIAS member.

\end{document}